\documentclass{article}

\usepackage{microtype}
\usepackage{graphicx}
\usepackage{subcaption}
\usepackage{booktabs}
\usepackage{multirow}
\usepackage{hyperref}
\usepackage{pifont}
\usepackage{xcolor}

\usepackage[preprint]{icml2026}

\usepackage{amsmath}
\usepackage{amssymb}

\newcommand{\cmark}{{\color{green!70!black}\ding{51}}}
\newcommand{\xmark}{{\color{red}\ding{55}}}

\graphicspath{{figures/}}

\icmltitlerunning{Effective Harness Engineering for Algorithm Discovery with Coding Agents}

\begin{document}

\twocolumn[
  \icmltitle{Effective Harness Engineering for Algorithm Discovery with Coding Agents}

  \icmlsetsymbol{equal}{*}

  \begin{icmlauthorlist}
    \icmlauthor{Yoichi Ishibashi}{nec}
    \icmlauthor{Taro Yano}{nec}
    \icmlauthor{Masafumi Oyamada}{nec}
  \end{icmlauthorlist}

  \icmlaffiliation{nec}{NEC Corporation}

  \icmlcorrespondingauthor{Yoichi Ishibashi}{yoichi-ishibashi@nec.com}

  \vskip 0.3in
]

\printAffiliationsAndNotice{}

\begin{abstract}
AlphaEvolve and FunSearch have demonstrated the potential of combining large language models (LLMs) with evolutionary search for automated algorithm discovery. However, discovery success is shaped not only by model capability but also significantly by the design of the execution infrastructure, i.e., the harness. This paper investigates effective harness design through three questions: under a fixed token budget, is it better to produce many algorithms with brief thought or fewer algorithms with deeper thought? How should the harness handle evaluation hacks, where generated programs exploit the scoring function? And how can agents that require full filesystem access execute safely in parallel? Using Vesper, an algorithm discovery framework that incorporates harness improvements addressing these questions, we evaluate on Circle Packing under the same token budget. Interestingly, generating fewer algorithms while thinking more deeply about each one achieved higher scores. That is, scaling the quality of each individual is more budget-efficient than scaling the number of evolutionary generations. Surprisingly, more capable models produced evaluation hacks at higher rates, making hack detection increasingly necessary as models scale.
\end{abstract}

\section{Introduction}

Scientific and engineering progress is often driven by the discovery of superior algorithms. Yet such algorithm design still relies heavily on manual effort by domain experts, and AI-driven automation of discovery is beginning to change this landscape~\cite{DBLP:journals/corr/abs-2505-13259}. In particular, the approach of using large language models (LLMs) as mutation operators in evolutionary search, iteratively creating algorithms that surpass human designs through cycles with an automatic evaluator, has achieved remarkable results. FunSearch evolved function-level algorithms with LLMs, generating mathematical constructions surpassing the best known results for the cap set problem~\cite{DBLP:journals/nature/RomeraParedesBNBKDREWFKF24}. Beyond improving algorithms alone, alternately improving both the generated algorithms and the generating model itself through reinforcement learning has also been explored~\cite{ishibashi2024can}. More recently, AlphaEvolve~\cite{DBLP:journals/corr/abs-2506-13131} expanded the scope of improvement to entire codebases and improved matrix multiplication algorithms that had remained unimproved for over 50 years. This technology is applicable to any domain where algorithms play a central role, including route optimization, chip placement design, and molecular generation in drug discovery, and may replace manual expert optimization.

What determines the success of algorithm discovery is not agent capability alone; the design of the agent infrastructure, i.e., the harness, has a substantial impact. The harness refers to the entire execution infrastructure that guides agents toward discovery, encompassing prompt construction, data design, evaluation pipelines, and parallel agent management. AlphaEvolve's success is supported not only by Gemini's model capability but also by the harness design as a whole, including randomized prompt switching, evaluation cascades, and an evolutionary database combining MAP-Elites with island-based population models.

Examining existing open-source implementations from this harness perspective reveals substantial room for improvement. Representative implementations such as OpenEvolve\footnote{\url{https://github.com/algorithmicsuperintelligence/openevolve}} and CodeEvolve~\cite{DBLP:journals/corr/abs-2510-14150} all use LLMs as stateless code generators through single-shot API calls, without leveraging the capabilities of coding agents that have rapidly advanced since 2025. Just as coding agents such as Claude Code\footnote{\url{https://github.com/anthropics/claude-code}} and Codex CLI\footnote{\url{https://github.com/openai/codex}} have transformed productivity in software development, differences in harness implementation may fundamentally change the practicality of algorithm discovery.

From these observations, we identify four practical challenges in existing harnesses. First, LLMs remain stateless code generators, without leveraging autonomous reasoning such as analyzing evaluation results, referencing the entire codebase, or planning multi-step corrections. Second, there is no mechanism to detect ``evaluation hacking''~\cite{DBLP:journals/corr/AmodeiOSCSM16}, where generated algorithms exploit flaws in the evaluation function rather than genuinely solving the problem. Third, running multiple search agents in parallel on a shared filesystem risks file conflicts and race conditions that compromise reproducibility. Fourth, each iteration starts from a blank slate, without systematically leveraging knowledge of successes and failures from past trials.

Beyond these challenges, a fundamental question about search strategy remains unanswered. Algorithm discovery is inherently a search problem whose success scales with the budget invested. The practical constraint is that every search step costs tokens, so token efficiency directly determines the scalability of discovery. Under a fixed token budget, is it better to produce many algorithms with brief thought, or fewer algorithms with deeper thought? Whether investing more thought per algorithm at the expense of fewer generations improves overall discovery performance requires empirical verification.

This paper proposes Vesper, a harness for LLM-driven algorithm discovery that integrates several candidate improvements, including coding agent integration (\autoref{sec:agent-integration}), evaluation hack detection (\autoref{sec:hack-detection}), and Git worktree isolation (\autoref{sec:worktree-isolation}), and empirically investigates which components matter and why. Interestingly, scaling the quality of each individual proved more budget-efficient than scaling the number of evolutionary generations (\autoref{sec:ablation}). We also find, surprisingly, that more capable frontier models such as GPT-5.2 are more prone to evaluation hacking.

\section{Related Work}

\paragraph{\textbf{LLM-Driven Algorithm Discovery}}
The dominant paradigm in LLM-driven algorithm discovery shares the common approach of incorporating LLMs into evolutionary operators as stateless generators that return single-shot code completions in response to prompts.
FunSearch~\cite{DBLP:journals/nature/RomeraParedesBNBKDREWFKF24} surpassed the best known results on the cap set problem, and AlphaEvolve~\cite{DBLP:journals/corr/abs-2506-13131} improved matrix multiplication algorithms that had remained unimproved for over 50 years.
Numerous specialized frameworks have also emerged, including EoH~\cite{DBLP:conf/icml/0044TY0LWL024}, ReEvo~\cite{DBLP:conf/nips/Ye0CBHKPS24}, LLaMEA~\cite{DBLP:journals/tec/SteinB25}, and AEL~\cite{DBLP:journals/corr/abs-2311-15249}. In open source, OpenEvolve reproduces AlphaEvolve's workflow, followed by CodeEvolve~\cite{DBLP:journals/corr/abs-2510-14150} and DeepEvolve~\cite{DBLP:journals/corr/abs-2510-06056}.

\paragraph{\textbf{Coding Agents}}
Since 2025, command-line coding agents that operate autonomously have emerged in rapid succession, including Codex CLI, Claude Code, Gemini CLI\footnote{\url{https://github.com/google-gemini/gemini-cli}}, and Qwen Code\footnote{\url{https://github.com/QwenLM/qwen-code}}.
These agents complete repository-wide reading, file editing, test execution, and debugging within a single session, with SWE-bench~\cite{DBLP:conf/iclr/JimenezYWYPPN24} driving their development as an evaluation platform.
Unlike single-shot API calls, coding agents can autonomously perform multi-step reasoning, including analyzing existing code, hypothesis testing based on execution results, and iterative error correction, giving them fundamentally different potential as evolutionary operators.
Vesper places the use of coding agents, rather than stateless API calls, at the center of its harness design.

\begin{figure*}[t!]
  \centering
  \includegraphics[width=\textwidth]{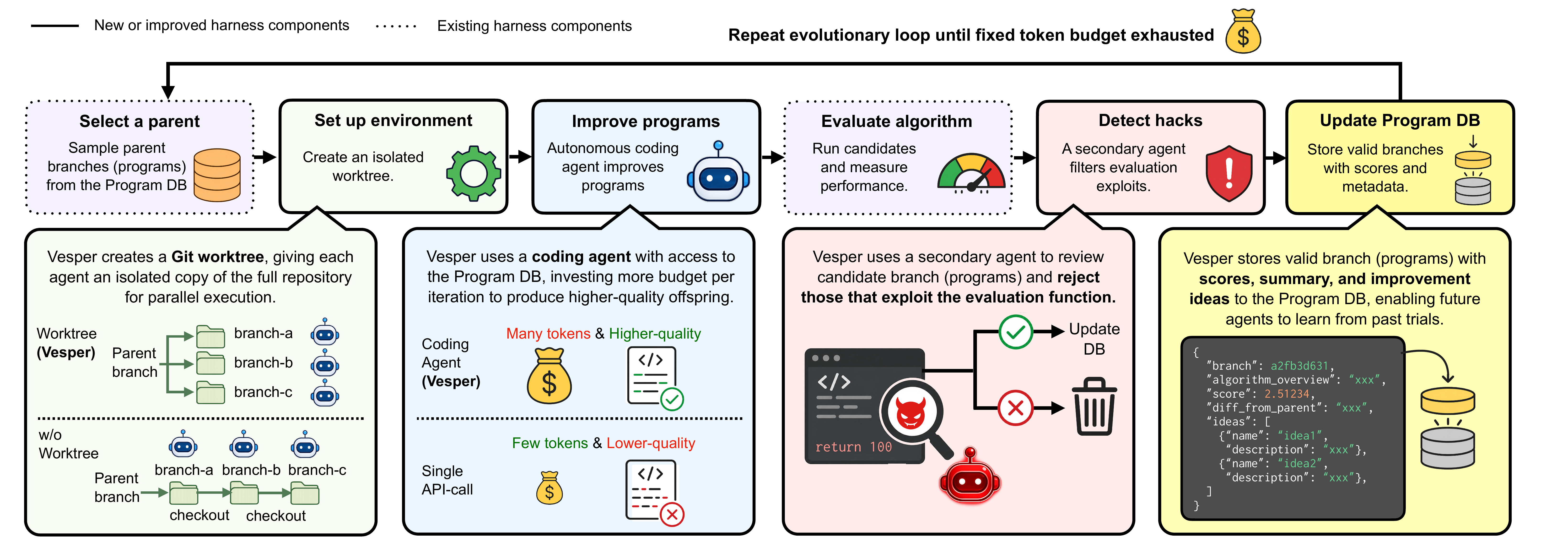}
  \caption{Vesper's evolutionary loop (top) and details of each harness improvement (bottom). Dotted borders indicate components shared with existing pipelines; solid borders indicate harness improvements introduced by Vesper. The following cycle repeats until the token budget is exhausted. (1)~Select a parent: sample a parent program (Git branch) from the program database. (2)~Set up environment: create a Git worktree from the parent branch, providing each agent with an isolated execution environment (\autoref{sec:worktree-isolation}). (3)~Improve programs: a coding agent autonomously improves code while referencing the program database (\autoref{sec:agent-integration}, \autoref{sec:experience}). (4)~Evaluate algorithm: compute the score of the improved algorithm. (5)~Detect hacks: a secondary agent reviews candidate programs and rejects those that exploit the evaluation function (\autoref{sec:hack-detection}). (6)~Update Program DB: store validated programs with scores, summaries, and improvement ideas in the program database. The accumulated trial history enables subsequent agents to analyze which approaches succeeded and which failed, informing their improvement decisions (\autoref{sec:experience}).}
  \label{fig:architecture}
\end{figure*}

\paragraph{\textbf{Agent Harnesses}}
LLM agent performance is shaped not only by the model but also substantially by the orchestration layer surrounding it, i.e., the harness. A harness refers to the entire runtime infrastructure including prompt construction, tool execution, context management, safety controls, and session persistence~\cite{pan2026naturallanguageagentharnesses}.
Recent empirical studies have shown that harness differences can dramatically change performance even with the same model. Meta-Harness~\cite{lee2026metaharnessendtoendoptimizationmodel} automatically searched over harness code while keeping model weights fixed, outperforming the best hand-designed harness by 7.7 points on text classification and achieving top rankings among agents using the same base model on TerminalBench-2.
In the coding agent domain as well, differences in harness-level design such as prompt design, tool selection strategies, and context management is more decisive than the choice of base model~\cite{bui2026buildingeffectiveaicoding}.
A similar structure is observed in algorithm discovery. AlphaEvolve's~\cite{DBLP:journals/corr/abs-2506-13131} success is supported not only by Gemini's model capability but by a harness design that includes evaluation cascades, randomized prompt switching, and an evolutionary database combining MAP-Elites with island-based population models.
This paper focuses on improving the harness.

\section{Overview of LLM-Driven Algorithm Discovery}
\label{sec:pipeline-overview}

\paragraph{\textbf{Background}}
To reproduce and extend these results, one must understand the common pipeline structure. This section provides an overview of the typical pipeline shared by AlphaEvolve and OpenEvolve.

\paragraph{\textbf{Algorithm Discovery Pipeline}}
A typical pipeline follows a loop structure consisting of five stages. First, \textbf{parent selection} chooses one or more parent programs from the population pool based on scores. Next, \textbf{prompt construction} assembles the selected parent's source code, problem description, and improvement instructions into an input prompt for the LLM. In the subsequent \textbf{LLM call} stage, the constructed prompt is sent to the LLM API, obtaining a code diff, i.e., a mutation, in a single API call. The resulting candidate is executed by an automatic evaluator in the \textbf{evaluation} stage, where a score is computed. Finally, in the \textbf{DB storage} stage, the score and program are saved to a database in preparation for the next cycle's parent selection. By repeating this cycle hundreds to thousands of times, increasingly high-scoring candidates accumulate from an initial seed program, ultimately leading to the discovery of algorithms surpassing human design.

\paragraph{\textbf{Diversity Maintenance via Island Model}}
Running this cycle with a single population can cause parent selection to converge prematurely around an early high-scoring solution, leaving other promising regions of the search space unexplored. To address this premature convergence, AlphaEvolve and OpenEvolve adopt the island model. The island model maintains $N$ independent population pools, running the pipeline independently on each island while periodically migrating top-scoring solutions between islands. This allows each island to pursue different search directions in parallel while propagating promising discoveries across the entire population, maintaining a balance between exploration and exploitation.

\section{Vesper}
\label{sec:vesper}

\subsection{Limitations of Existing Pipelines and Design Principles}

\paragraph{\textbf{Limitations of Existing Pipelines}}
The pipeline overviewed in \autoref{sec:pipeline-overview} has four areas open for improvement from a harness perspective. First, LLM calls are single-shot and stateless, preventing multi-step reasoning such as diagnosing the cause of a low score and re-debugging the code. Second, there is no verification mechanism after the evaluation stage, allowing hack solutions that exploit evaluation function flaws to be stored in the database and contaminate parent selection. Third, when running multiple LLM calls in parallel, conflicts on a shared filesystem can compromise reproducibility and stability. Fourth, each iteration starts from a blank slate without systematic utilization of knowledge about which approaches succeeded and which failed in past trials.

\paragraph{\textbf{Vesper's Architecture}}
Vesper integrates improvements addressing each of these four limitations into a single evolutionary loop (\autoref{fig:architecture}). Vesper treats the target of evolution as a repository, building an evolutionary tree with Git branches as units. In the initial state, only a single seed branch containing the algorithm and evaluation code is registered in the program database. In each cycle, a parent branch is first sampled from the program database, and a Git worktree is created from that branch to construct an isolated execution environment (\autoref{sec:worktree-isolation}). A coding agent is then launched within the worktree, where it references the program database to analyze past success and failure patterns (\autoref{sec:experience}) while autonomously performing code improvement and testing (\autoref{sec:agent-integration}). After completion, the evaluator computes a score, and a hack detection step verifies candidate integrity (\autoref{sec:hack-detection}). Candidate branches passing verification are added to the program database. Vesper maintains diversity solely through the island model, without the MAP-Elites component used in AlphaEvolve's database~\cite{DBLP:journals/corr/MouretC15}.

\subsection{Coding Agent Integration}
\label{sec:agent-integration}

\paragraph{\textbf{Limitations of Stateless Generation}}
In representative algorithm discovery pipelines such as OpenEvolve, a prompt is sent to the LLM API and a code diff is received as a single response. While it is possible to include previous evaluation results and error information in the next iteration's prompt, that constitutes a separate iteration; achieving a multi-step debugging loop where code is executed, errors are observed, and corrections are attempted within the same session requires harness modifications. For example, if a runtime exception occurs in generated code, it is not possible to inspect the error and apply repeated fixes on the spot; a single failed generation wastes the entire iteration.

\paragraph{\textbf{Quality vs.\ Quantity Trade-off}}
Evolutionary search admits two strategies: spend fewer tokens per iteration to generate many candidates and rely on generational turnover, or invest more tokens per iteration to produce fewer but higher-quality candidates. Stateless API calls correspond to the former: they are cheap but lack self-correction, placing an upper bound on candidate quality. Vesper adopts the latter strategy by replacing this step with autonomous coding agents. Agents consume several times more tokens per algorithm due to repository reading, multi-step reasoning, test execution, and debugging, but substantially improve candidate quality through self-correction within a single session. This quality-versus-quantity trade-off is the central question of this paper and is empirically examined in \autoref{sec:ablation}.

\paragraph{\textbf{Autonomous Operation and Structured Output}}
Each agent operates as an autonomous coding session. It reads source files, runs evaluations, inspects errors, and references related modules on its own, submitting candidate solutions through multiple tool-use turns. The improvement strategy, including which functions to modify and how, is determined by the agent itself, free from the single-function mutation constraints typical of prompt-based systems. Agent outputs are recorded in a structured format containing not only code changes but also descriptions of attempted approaches and rationale for score improvements, serving as input to the DB observation mechanism (\autoref{sec:experience}).

\subsection{Evaluation Hack Detection}
\label{sec:hack-detection}

\paragraph{\textbf{Risk of Evaluation Hacking}}
Reward hacking, where agents exploit the reward function itself rather than the intended task, is widely known in reinforcement learning~\cite{DBLP:journals/corr/AmodeiOSCSM16, DBLP:conf/nips/SkalseHKK22}. In LLM-driven algorithm discovery, there is a risk that generated programs achieve high scores through hardcoding expected outputs or exploiting boundary conditions of the scoring function, and this risk becomes particularly acute when agents are powerful enough to inspect and modify the evaluation infrastructure. In the evolutionary setting, if a single hack solution with an inflated score dominates parent selection, degenerate strategies propagate throughout the population, rendering subsequent search effectively meaningless. No existing framework incorporates a mechanism to detect this problem.

\paragraph{\textbf{Hack Detection Mechanism}}
Vesper addresses this problem by executing a secondary agent-based verification pass after each candidate passes evaluation. An independent agent session inspects the candidate's implementation code and determines whether the solution genuinely addresses the algorithmic problem or exploits the evaluation mechanism. Candidates flagged as hacks are excluded from the parent selection pool, preventing degenerate solutions from spreading through the evolutionary lineage.

\subsection{DB Observation}
\label{sec:experience}

\paragraph{\textbf{Inefficiency of Stateless Search}}
In existing pipelines, each iteration effectively starts from a blank slate. The prompt contains only the parent code, without systematic information about which approaches were previously tried, what succeeded, and what failed. Consequently, search efficiency is substantially degraded by retrying already-failed directions or failing to leverage promising approaches discovered in other search threads.

\paragraph{\textbf{Program Database}}
Vesper addresses this problem through an SQLite-based program database. Repository information for each worktree, branch lineage, evaluation results, algorithm descriptions, code diffs, and improvement ideas are accumulated in relational tables. By aggregating the structured outputs recorded by each agent into this database, individual trials become shared knowledge assets rather than disposable artifacts.

\paragraph{\textbf{Autonomous DB Observation by Agents}}
Accumulating experience is meaningless unless it is properly delivered to agents. Injecting summaries into prompts requires the harness to predetermine which information is useful for the current improvement, precluding flexible, situation-dependent knowledge retrieval. Vesper instead passes the database path to agents, allowing them to execute SQL queries themselves to retrieve needed information. Since coding agents have tool execution capabilities, they can, for example, trace their parent branch's lineage to avoid past failures or investigate the changes in algorithms that produced rapid score increases on other islands to inform their own improvements. The division of roles between DB observation and island model migration is discussed in \autoref{app:db-migration}.

\subsection{Git Worktree Isolation}
\label{sec:worktree-isolation}

\paragraph{\textbf{Conflicts in Parallel Execution}}
When multiple agents simultaneously write to the same files, race conditions and state corruption occur. This risk is particularly severe because Vesper's agents have unrestricted filesystem access within their workspace. Traditional approaches can achieve isolation through full repository cloning, but this is impractical for large target repositories in terms of disk consumption and initialization time.

\paragraph{\textbf{Isolation via Worktrees}}
Vesper assigns each agent a dedicated Git worktree, achieving complete filesystem isolation without cloning the entire repository. Worktrees are a native Git feature that provides independent working directories while sharing the repository's internal data. This allows multiple agents to operate safely in parallel without conflicts. The parallelism efficiency achieved by this isolation is evaluated in \autoref{sec:ablation}.

\section{Harness Comparison Experiments}
\label{sec:ablation}

\begin{table*}[t]
  \caption{Harness comparison on Circle Packing ($n{=}26$, 2 runs). Model refers to the foundation model of the agent used for algorithm discovery. Hack detection uses \texttt{gpt-5.1-codex-mini} for all conditions. Token Budget is the total token allocation for each experiment. Raw Best is the highest score recognized by the system (hack solutions are already excluded when hack detection is enabled). Best additionally applies mechanical exclusion of scores $>3$ for $\dagger$ conditions, reporting the highest valid score across 2 runs (sum of radii, higher is better). \#Algo, Tok/Algo, Cost(\$), Hacks, and Hack\% are averaged over 2 runs. $\dagger$ indicates conditions without hack detection, where scores $>3$ were mechanically excluded before reporting the best.}
  \label{tab:experiments}
  \centering
  \fontsize{8}{9.5}\selectfont
  \setlength{\tabcolsep}{3pt}
  \begin{tabular}{@{}llccccrr|rcrrc@{}}
    \toprule
    Harness & Model & Agent & Hack & DB & Raw Best & Best & Token Budget & \#Algo & Tok/Algo & Cost(\$) & Hacks & Hack\% \\
    \midrule
    Vesper     & \texttt{5.2-codex}      & \cmark & \cmark & \cmark & 2.63110 & 2.63110 & 40M & 568 & 70.5K & 391 & 92 & 16.6\% \\
    Vesper     & \texttt{5.2-codex}      & \cmark & \cmark & \xmark & 2.63599 & \textbf{2.63599} & 40M & 338 & 118.8K & 391 & 26 & 7.8\% \\
    Vesper     & \texttt{5.2-codex}      & \cmark & \xmark & \cmark & $>$10$^{10}$ & \textbf{2.63599}$^\dagger$ & 40M & 742 & 54.2K & 391 & --- & --- \\
    Vesper     & \texttt{5.2-codex}      & \cmark & \xmark & \xmark & $>$10$^{10}$ & \textbf{2.63599}$^\dagger$ & 40M & 452 & 89.6K & 391 & --- & --- \\
    \midrule
    Vesper     & \texttt{5.1-codex-mini} & \cmark & \cmark & \cmark & 2.61232 & 2.61232 & 40M & 87 & 465.2K & 42 & 0 & 0\% \\
    Vesper     & \texttt{5.1-codex-mini} & \cmark & \cmark & \xmark & 2.62721 & 2.62721 & 40M & 101 & 400.1K & 42 & 0 & 0\% \\
    Vesper     & \texttt{5.1-codex-mini} & \cmark & \xmark & \cmark & 2.63598 & 2.63598$^\dagger$ & 40M & 90 & 451.7K & 42 & --- & --- \\
    Vesper     & \texttt{5.1-codex-mini} & \cmark & \xmark & \xmark & 2.63586 & 2.63586$^\dagger$ & 40M & 110 & 369.6K & 42 & --- & --- \\
    \midrule
    OpenEvolve & \texttt{5.2}             & \xmark & \xmark & \xmark & 2.41852 & 2.41852$^\dagger$ & 40M & 1{,}671 & 23.9K & 107 & --- & --- \\
    OpenEvolve & \texttt{5.2-codex}      & \xmark & \xmark & \xmark & 2.54142 & 2.54142$^\dagger$ & 40M & 1{,}510 & 26.5K & 245 & --- & --- \\
    OpenEvolve & \texttt{5.1-codex-mini}  & \xmark & \xmark & \xmark & 2.48092 & 2.48092$^\dagger$ & 40M & 1{,}487 & 26.9K & 27 & --- & --- \\
    \midrule
    AlphaEvolve & Gemini & \xmark & \xmark & \xmark & --- & 2.6358 & --- & --- & --- & --- & --- & --- \\
    Human best & --- & --- & --- & --- & --- & 2.6340 & --- & --- & --- & --- & --- & --- \\
    \bottomrule
  \end{tabular}
\end{table*}

\subsection{Contribution of Each Component to Discovery Performance}

\paragraph{\textbf{Experimental Objective}}
We seek to quantitatively verify the extent to which the harness improvements proposed in \autoref{sec:vesper} affect algorithm discovery performance. If Vesper outperforms the baseline under the same model and token budget, this provides evidence that harness design influences discovery performance. Furthermore, by controlling the presence or absence of hack detection and DB observation, we isolate the contribution of each component.

\paragraph{\textbf{Baseline Selection}}
To examine how harness design differences affect algorithm discovery performance, we compare against OpenEvolve\footnote{v0.2.25}, an unofficial open-source reimplementation of AlphaEvolve, as our baseline. OpenEvolve is a harness designed with prompt construction and evaluation pipelines built around stateless API calls. Vesper is not a modification of OpenEvolve but a system built from the ground up with coding agents as the design premise, resulting in four key differences: coding agent integration, evaluation hack detection, Git worktree isolation, and DB observation for leveraging past experience. Simply plugging an agent backend into OpenEvolve would not leverage agent-specific capabilities such as autonomous codebase access and multi-step reasoning, so we compare the overall harness design rather than swapping individual components.

\paragraph{\textbf{Experimental Conditions}}
We use Circle Packing ($n{=}26$) as the task: a geometric optimization problem of placing 26 non-overlapping circles in a unit square to maximize the sum of radii. This problem has been adopted as a standard benchmark across many LLM-driven algorithm discovery studies~\cite{DBLP:journals/corr/abs-2506-13131,DBLP:journals/corr/abs-2511-23473,DBLP:journals/corr/abs-2511-08522}. We target AlphaEvolve's achievement of 2.635. To enable fair comparison between systems with different per-iteration costs, we adopt a token consumption ceiling (40M tokens) rather than number of generated algorithms as the termination criterion. Token budget control directly maps to API cost constraints in practical deployment, making it a more pragmatic evaluation criterion. Island model hyperparameters (5 islands, migration interval 50, migration rate 0.1, exploration ratio 0.3, exploitation ratio 0.7) follow OpenEvolve's official settings and are unified across both systems. OpenEvolve uses \texttt{gpt-5.2} as its default model. Codex-series models offer stronger code-specialized reasoning capabilities than Chat-series models but use a different API endpoint incompatible with OpenEvolve's Chat Completions interface. To avoid conflating harness effects with model differences, we adapted OpenEvolve to support Codex-series models and evaluate both systems under the same model, isolating the contribution of harness design. Note that OpenEvolve invokes these models through single-shot API calls, without multi-step reasoning or codebase access, so the same model operates under fundamentally different usage patterns across the two harnesses. The comparison factors are harness (Vesper / OpenEvolve), agent usage, model (expensive \texttt{gpt-5.2-codex}\footnote{\url{https://platform.openai.com/docs/models/gpt-5.2-codex}} / inexpensive \texttt{gpt-5.1-codex-mini}\footnote{\url{https://platform.openai.com/docs/models/gpt-5.1-codex-mini}}), hack detection presence, and DB observation (the feature where agents autonomously reference past trial histories via an SQLite database, \autoref{sec:experience}) presence, yielding the conditions shown in \autoref{tab:experiments}.

\begin{figure*}[t]
  \centering
  \includegraphics[width=\textwidth]{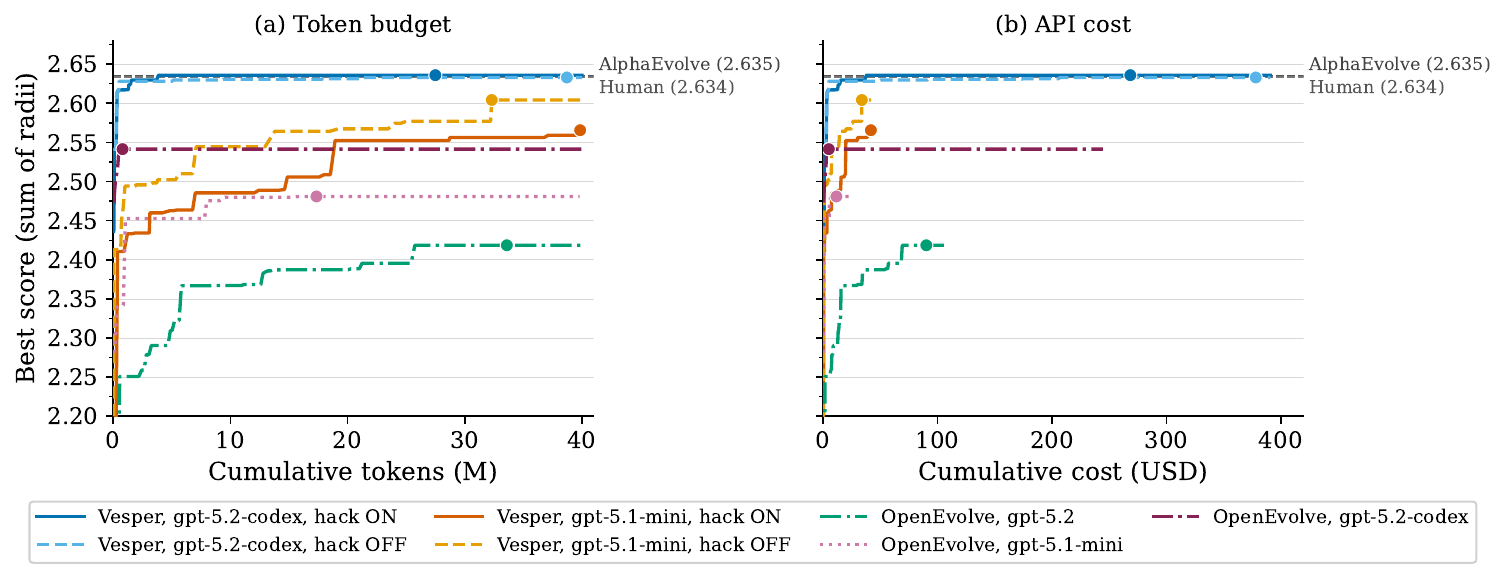}
  \caption{Best score progression. (a)~Cumulative tokens, (b)~cumulative API cost. Each marker represents an individual (algorithm) that updated the best score. Conditions without DB observation only. Algorithms flagged as evaluation hacks are excluded. Dashed lines indicate AlphaEvolve (2.635) and the human best (2.634).}
  \label{fig:best-score}
\end{figure*}

\paragraph{\textbf{Finding 1: Replacing with Coding Agents Substantially Improves Discovery Performance}}
In LLM-driven algorithm discovery, model capability tends to attract attention, but how much do results change when the harness differs under the same model and token budget? To exclude the contribution of hack detection, we compare conditions without hack detection. From \autoref{tab:experiments}, Vesper (\texttt{gpt-5.2-codex}, no hack detection) substantially outperforms OpenEvolve. As shown in \autoref{fig:best-score}(a), Vesper rapidly improves its score from the early stages of search, already surpassing OpenEvolve's final score at approximately 5M tokens. This effect is independent of model scale: even the inexpensive \texttt{gpt-5.1-codex-mini} (no hack detection) outperforms OpenEvolve's \texttt{gpt-5.2}. The harness-level differences between Vesper and OpenEvolve, including coding agent integration, evaluation hack detection, and Git worktree isolation, collectively produce the performance gap. Moreover, OpenEvolve with \texttt{gpt-5.1-codex-mini} confirms that even under the same model, harness design alone produces a substantial performance gap.

\begin{figure}[t]
  \centering
  \includegraphics[width=\columnwidth]{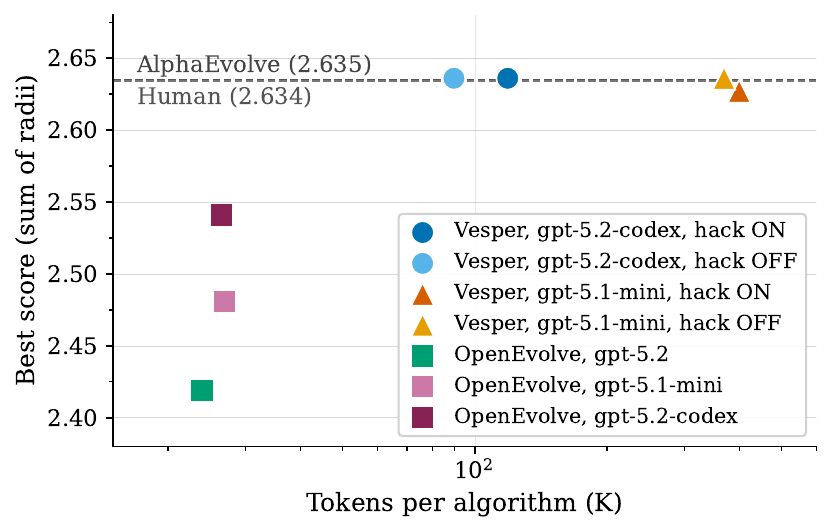}
  \caption{Relationship between per-iteration token investment (Tok/Algo) and best score. Each marker represents an experimental condition from \autoref{tab:experiments}. Higher compute investment per iteration yields higher best scores. Conditions without DB observation only.}
  \label{fig:efficiency-frontier}
\end{figure}

\paragraph{\textbf{Finding 2: Scaling Reasoning per Algorithm Is More Efficient than Scaling Generations}}
Under the same token budget, is it better to generate many low-quality candidates or fewer high-quality ones? The Tok/Algo column in \autoref{tab:experiments} answers this question. OpenEvolve (\texttt{gpt-5.2}) spends 23.9K tokens per algorithm and generates 1,671 candidates within 40M tokens, yet its best score remains low. In contrast, Vesper (\texttt{gpt-5.2-codex}, no hack detection) spends 89.6K tokens per algorithm and generates only 452 candidates, yet surpasses both AlphaEvolve's achievement and the human best. \autoref{fig:best-score}(a) makes this gap even more vivid: Vesper surpasses OpenEvolve's final score at approximately 5M tokens, a budget in which OpenEvolve generates over 200 candidates. The compute-intensive processes of coding agents, including multi-step reasoning, codebase inspection, and evaluation result analysis, dramatically elevate the quality of each candidate. This trend holds for inexpensive models as well: Vesper (\texttt{gpt-5.1-codex-mini}) with 110 candidates outperforms OpenEvolve's 1,487 candidates and approaches the human best. Producing fewer high-quality algorithms is more efficient than producing many low-quality ones. That is, under a fixed budget, scaling reasoning per algorithm is more effective than scaling the number of generations. \autoref{fig:efficiency-frontier} summarizes this relationship: conditions with higher per-iteration token investment (Tok/Algo) consistently achieve higher best scores, with a clear separation between OpenEvolve (lower left) and Vesper (upper right). A detailed cost comparison is provided in \autoref{sec:cost-performance}.

\paragraph{\textbf{Finding 3: More Capable Models Generate More Evaluation Hacks}}
By comparing conditions that differ only in the presence of hack detection within the same harness, we isolate the contribution of evaluation hack detection. For \texttt{gpt-5.2-codex}, hack detection on outperforms off. As shown in \autoref{tab:experiments}, 29 out of 352 algorithms (8.2\%) were detected and excluded as evaluation hacks under this condition. By eliminating hack solutions from the population pool, subsequent parent selection operates exclusively on sound solutions, maintaining search quality. In contrast, for \texttt{gpt-5.1-codex-mini}, hack detection on underperforms off. Under this condition, no evaluation hacks occurred at all (\autoref{tab:experiments}). More capable models have greater ability to generate code exploiting evaluation function vulnerabilities, and the necessity of hack detection increases in proportion to model capability. When no hacks occur, the overhead of hack detection reduces the number of generations under the same budget, resulting in lower scores.

\paragraph{\textbf{Finding 4: The Effect of DB Observation Is Limited}}
DB observation is a feature that allows agents to reference past trial histories via an SQLite database, enabling each agent to leverage past successes and failures when deciding improvement directions. However, the two-run results in \autoref{tab:experiments} show no clear benefit. For \texttt{gpt-5.1-codex-mini} (hack detection on), DB observation on underperforms off, and for \texttt{gpt-5.2-codex} (hack detection on), DB observation on likewise underperforms off. DB observation consumes tokens, reducing the number of generated algorithms under the same budget, and this loss of search opportunities may offset the benefit of referencing past trials.

\subsection{Parallel Search via Worktree Isolation}

\paragraph{\textbf{Worktree Isolation Enables Efficient Parallel Search}}
OpenEvolve generates complete programs as strings via stateless API calls and writes them to temporary files only during evaluation, requiring no filesystem isolation for parallelism. In contrast, Vesper's agents directly read, modify, and test the shared codebase, making concurrent access without isolation impossible. Git worktrees provide independent working directories while sharing the repository's internal data, enabling safe parallel execution. \autoref{tab:parallelism} shows the effective parallelism across all conditions with 4 parallel agents. The parallelism ratio (total sequential agent time divided by wall-clock time) ranges from 3.2x to 3.9x, reducing wall-clock time from approximately 70 hours to 20 hours in the most compute-intensive condition.

\begin{table}[t]
  \caption{Effect of worktree isolation on execution time. Without worktree isolation (WT \xmark), a single agent runs one branch at a time; with worktree isolation (WT \cmark), 4 agents run in parallel on independent worktrees. $^\dagger$ Estimated from the sum of per-agent execution times.}
  \label{tab:parallelism}
  \centering
  \fontsize{8}{9.5}\selectfont
  \setlength{\tabcolsep}{4pt}
  \begin{tabular}{@{}llcccrr@{}}
    \toprule
    Model & Hack & DB & WT & Exec.\ Time (h) & Speedup \\
    \midrule
    \multirow{2}{*}{\texttt{5.2-codex}} & \multirow{2}{*}{\cmark} & \multirow{2}{*}{\cmark} & \xmark & 46.6$^\dagger$ & \multirow{2}{*}{3.2x} \\
    & & & \cmark & \textbf{14.5} & \\
    \midrule
    \multirow{2}{*}{\texttt{5.2-codex}} & \multirow{2}{*}{\cmark} & \multirow{2}{*}{\xmark} & \xmark & 69.7$^\dagger$ & \multirow{2}{*}{3.5x} \\
    & & & \cmark & \textbf{20.0} & \\
    \midrule
    \multirow{2}{*}{\texttt{5.2-codex}} & \multirow{2}{*}{\xmark} & \multirow{2}{*}{\cmark} & \xmark & 40.5$^\dagger$ & \multirow{2}{*}{3.9x} \\
    & & & \cmark & \textbf{10.4} & \\
    \midrule
    \multirow{2}{*}{\texttt{5.2-codex}} & \multirow{2}{*}{\xmark} & \multirow{2}{*}{\xmark} & \xmark & 64.3$^\dagger$ & \multirow{2}{*}{3.8x} \\
    & & & \cmark & \textbf{17.1} & \\
    \midrule
    \multirow{2}{*}{\texttt{5.1-codex-mini}} & \multirow{2}{*}{\cmark} & \multirow{2}{*}{\cmark} & \xmark & 13.9$^\dagger$ & \multirow{2}{*}{3.3x} \\
    & & & \cmark & \textbf{4.2} & \\
    \midrule
    \multirow{2}{*}{\texttt{5.1-codex-mini}} & \multirow{2}{*}{\cmark} & \multirow{2}{*}{\xmark} & \xmark & 16.1$^\dagger$ & \multirow{2}{*}{3.3x} \\
    & & & \cmark & \textbf{4.9} & \\
    \midrule
    \multirow{2}{*}{\texttt{5.1-codex-mini}} & \multirow{2}{*}{\xmark} & \multirow{2}{*}{\cmark} & \xmark & 15.2$^\dagger$ & \multirow{2}{*}{3.6x} \\
    & & & \cmark & \textbf{4.2} & \\
    \midrule
    \multirow{2}{*}{\texttt{5.1-codex-mini}} & \multirow{2}{*}{\xmark} & \multirow{2}{*}{\xmark} & \xmark & 30.6$^\dagger$ & \multirow{2}{*}{3.3x} \\
    & & & \cmark & \textbf{9.3} & \\
    \bottomrule
  \end{tabular}
\end{table}

\section{Cost-Performance Analysis}
\label{sec:cost-performance}

\paragraph{\textbf{Objective of Analysis}}
\autoref{sec:ablation} showed that fewer high-quality candidates outperform many low-quality ones under the same token budget. However, coding agents use Codex CLI models with higher per-token costs than stateless calls (\$9.77/M tokens vs \$2.68/M tokens), creating an approximately 3.7$\times$ gap in actual cost under the same token budget. Whether the quality-focused strategy remains superior even after accounting for this cost difference is an important practical question. Furthermore, it remains unclear whether one should select an expensive model for the agent backend or use an inexpensive model to maximize iterations. This section answers these two questions.

\paragraph{\textbf{Experimental Setup}}
In addition to the Vesper (\texttt{gpt-5.2-codex}) and OpenEvolve results from \autoref{sec:ablation}, we conduct a supplementary experiment increasing OpenEvolve's token budget to 146M tokens, equivalent to Vesper's actual cost (approximately \$392), enabling comparison under equal expenditure. For model selection comparison, we use four Vesper conditions on the same harness with \texttt{gpt-5.2-codex} and \texttt{gpt-5.1-codex-mini}.

\paragraph{\textbf{Finding 1: The Quality-Focused Strategy Remains Superior Even at Equal Cost}}
\autoref{sec:ablation} compared systems under the same token budget, but what happens when actual cost is equalized? We increased OpenEvolve's token budget to 146M tokens, equivalent to Vesper's cost (approximately \$392) (\autoref{app:cost-comparison}). OpenEvolve generated 4,239 algorithms but failed to close the gap with Vesper (\autoref{tab:experiments}). As shown in \autoref{fig:best-score}(b), Vesper already far exceeds OpenEvolve's final level at approximately \$107 worth (11M tokens), and OpenEvolve fails to reach this level even at equivalent expenditure. Even after accounting for per-token cost differences, the quality-focused strategy of investing more compute per iteration remains superior.

\paragraph{\textbf{Finding 2: Expensive Models Offer Better Cost-Performance than Inexpensive Models}}
From \autoref{fig:best-score}(b), the \texttt{gpt-5.2-codex} curve consistently lies above the \texttt{gpt-5.1-codex-mini} curve on the cost axis. That is, for the same monetary investment, the expensive model reaches a higher score. Although \texttt{gpt-5.2-codex} has a higher per-token cost (\$9.77/M tokens vs \$1.05/M tokens) and therefore generates fewer algorithms under the same budget, the quality per algorithm is higher, maintaining superiority in performance improvement per dollar. When budget permits, selecting the expensive model is also rational from a cost-effectiveness perspective.

\paragraph{\textbf{Finding 3: \$38 Reaches AlphaEvolve-Level Performance}}
\texttt{gpt-5.2-codex} reaches the human best and AlphaEvolve-level performance at approximately \$38 (3.9M tokens) (\autoref{fig:best-score}(b)). Meanwhile, the inexpensive \texttt{gpt-5.1-codex-mini} approaches the human best at a total cost of \$42. Even the inexpensive model approaches the human best at \$42.

\section{Conclusion}
\label{sec:conclusion}

This paper proposed Vesper, a framework that systematically strengthens the harness for LLM-driven algorithm discovery. In Circle Packing ($n{=}26$) experiments, Vesper substantially outperformed OpenEvolve under the same token budget, surpassing both AlphaEvolve's achievement and the human best. Under a fixed budget, investing more tokens per iteration to produce fewer high-quality candidates via coding agents proved more efficient than generating many low-quality candidates through cheap API calls, and this advantage held even in cost-based comparisons accounting for per-token price differences. Furthermore, more capable models generate more evaluation hacks, demonstrating that the necessity of hack detection increases with model capability. These results demonstrate that LLM-driven algorithm discovery performance is substantially influenced not only by model capability but also by harness design, and Vesper's design provides concrete guidelines for making this field more practical.

\bibliography{references}
\bibliographystyle{icml2026}

\clearpage
\appendix
\section{Detailed Experimental Settings}
\label{app:experiment-details}

The detailed settings used in the experiments of \autoref{sec:ablation} and \autoref{sec:cost-performance} are provided below.

\paragraph{\textbf{Island Model Settings}}
Island model hyperparameters follow OpenEvolve's official settings and are unified across both Vesper and OpenEvolve. \autoref{tab:island-params} shows the parameter values. Parent selection uses a three-way probability mixture: with probability 0.3 for exploration (sampling from the bottom 80\% by score), probability 0.7 for exploitation (sampling from the top 20\% by score), and the remaining probability 0.0 for uniform random selection. Migration occurs in a ring topology every 50 completed algorithms, with the top 10\% of each island copied to adjacent islands.

\begin{table}[t]
  \caption{Island model hyperparameters. Unified across both systems.}
  \label{tab:island-params}
  \centering
  \small
  \begin{tabular}{@{}lr@{}}
    \toprule
    Parameter & Value \\
    \midrule
    Number of islands & 5 \\
    Migration interval & 50 algorithms \\
    Migration rate & 0.1 \\
    Exploration ratio $p_{\mathrm{explore}}$ & 0.3 \\
    Exploitation ratio $p_{\mathrm{exploit}}$ & 0.7 \\
    Max population per island & 5 \\
    Max total population & 25 \\
    \bottomrule
  \end{tabular}
\end{table}

\paragraph{\textbf{Vesper Agent Settings}}
Vesper runs up to 4 agents in parallel, with each agent operating within a dedicated Git worktree. The agent timeout is set to 1 hour. For conditions where evaluation hack detection is enabled, the detection agent uses the inexpensive \texttt{gpt-5.1-codex-mini}, performing hack detection with a model separate from the main agent.

\paragraph{\textbf{OpenEvolve Settings}}
OpenEvolve calls the LLM statelessly through the Chat Completions API. LLM temperature is set to 0.7, top\_p to 0.95, and maximum output tokens to 32,768. Prompts include the top 3 programs by score and 2 programs for diversity, with template stochasticity enabled. Evaluation uses cascade evaluation (thresholds 0.5, 0.75) with a maximum of 4 parallel executions. The random seed is fixed at 42.

\paragraph{\textbf{Cost Estimation Method}}
To estimate costs from token counts, we calculated blended rates (mixed unit prices) per model from OpenAI's billing data. The blended rate is the total cost divided by total tokens for each model, representing a weighted average of input, cached input, and output pricing. \texttt{gpt-5.2-codex} is \$9.77/M tokens, \texttt{gpt-5.1-codex-mini} is \$1.05/M tokens, and \texttt{gpt-5.2} (API) is \$2.68/M tokens. Cumulative cost for each experiment is estimated by multiplying cumulative tokens by the blended rate.

\section{Division of Roles Between DB Observation and Island Model Migration}
\label{app:db-migration}

DB observation and island model migration play complementary roles. Migration directly adds top solutions from other islands into the local parent selection pool, sharing the solutions themselves between islands. DB observation, in contrast, enables agents to reference the trial history across all islands, acquiring strategy-level knowledge about which approaches were effective and which failed. Even when an agent discovers a high-scoring solution from another island in the DB, it serves only as reference information and does not enter the parent selection candidates. Migration handles direct propagation of solutions, while DB observation reinforces improvement direction decisions; this division allows both to function independently.

\section{Cost Comparison at Equal Expenditure}
\label{app:cost-comparison}

\begin{figure*}[t]
  \centering
  \includegraphics[width=\textwidth]{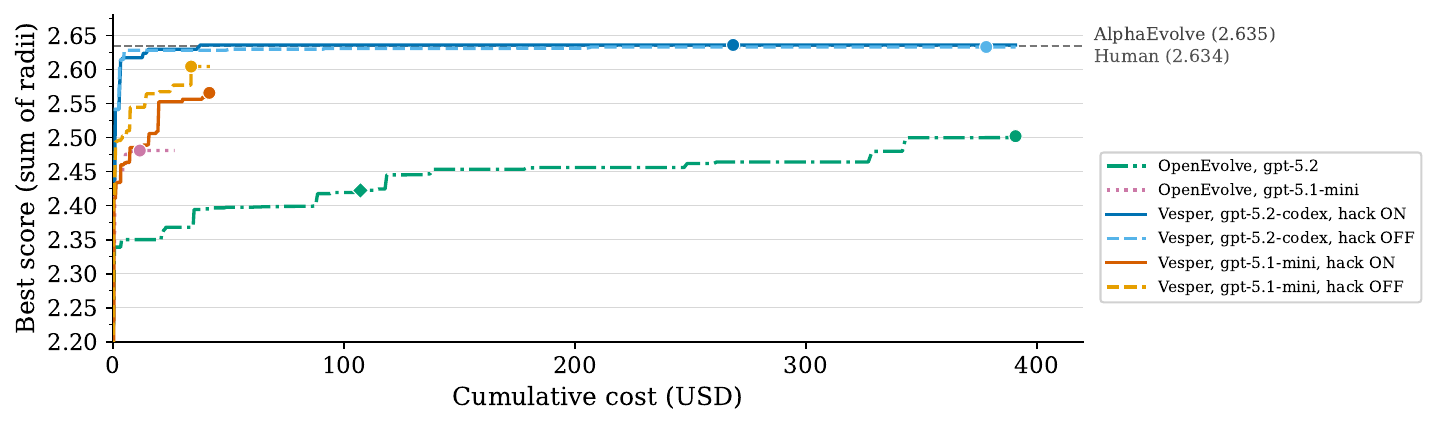}
  \caption{Best score progression versus cumulative API cost. Each marker represents an individual (algorithm) that updated the best score. Even when OpenEvolve's budget is expanded to match Vesper's cost (\$392, 146M tokens), its score plateaus at 2.502. Vesper approaches the human best at \$42 (\texttt{gpt-5.1-codex-mini}) and surpasses AlphaEvolve at \$391 (\texttt{gpt-5.2-codex}). Diamonds indicate OpenEvolve at the 40M token (\$107) point.}
  \label{fig:cost-comparison}
\end{figure*}

\autoref{fig:cost-comparison} shows the best score progression versus cumulative API cost, including a supplementary experiment where OpenEvolve's token budget was expanded to match Vesper's cost (\$392, 146M tokens).

\end{document}